\documentstyle[prb,aps,psfig]{revtex}
\begin{document}

\title{Dynamic Density Functional Theory of Fluids}

\author{Umberto Marini Bettolo Marconi$^a$ and
Pedro Tarazona$^b$}  

\address
{ $^a$ Dipartimento di Matematica e Fisica and Istituto Nazionale
di Fisica della Materia,
Universit\`a di Camerino, \\
Via Madonna delle Carceri, 62032, Camerino (Italy)  \\
e-mail: umberto.marini.bettolo@roma1.infn.it\\
$^b$ Departamento de F\'{\i}sica Te{\'o}rica de la Materia Condensada, 
Universidad Aut{\'o}noma de Madrid, 28049 Madrid (Spain)}

\date{\today}
\maketitle

\begin{abstract}
 We present a new time-dependent Density Functional 
approach to study the relaxational
dynamics of an assembly of interacting particles subject to thermal noise.
Starting from the Langevin stochastic equations of motion for
the velocities of the particles we are able by means of an
approximated closure to derive a self-consistent deterministic equation 
for the temporal evolution of the average particle density.
 The closure is equivalent to assuming that the equal-time two-point 
correlation function out of equilibrium has the same properties as 
its equilibrium version. 
 The changes in time of the density depend on the functional derivatives
of the grand canonical free energy functional $F[\rho]$ of the system.
In particular the static solutions of the  equation for the density correspond
to the exact equilibrium profiles provided one is able to determine the
exact form of $F[\rho]$. 
In order to assess the validity of our approach we performed a comparison
between the Langevin dynamics and the dynamic density functional method
for a one-dimensional hard-rod system in three relevant cases and
found remarkable agreement, with some interesting exceptions, which are
discussed and explained. In addition, we consider the case where one
is forced to use an approximate form of $F[\rho]$.
 Finally we compare the present method with the stochastic
equation for the density proposed
by other authors [Kawasaki,Kirkpatrick etc.] and discuss
the role of the thermal fluctuations.

\end{abstract} 
\newpage
\section{Introduction}

In recent years the off-equilibrium properties of extended systems
have represented a very active field of research. In fact, while the
present understanding of systems in thermodynamic 
equilibrium is rather satisfactory
and is based on well established theoretical methods, the comprehension
of their dynamical aspects is far from complete,
in spite of massive experimental and theoretical investigations
\cite{Bray,Gunton}.

In the present paper we shall focus attention on the 
Density Functional method, which
represents a powerful and widely used tool to investigate the
static properties of many particles systems \cite{Evans}
and consider the possibility of extending this approach to off-equilibrium
situations.  
Some authors have already employed similar approaches on a purely 
phenomenological basis by analogy with the popular Ginzburg-Landau
time dependent equation and the Cahn-Hilliard equation, but these 
methods are not applicable to the highly structured density profiles
that one observes at the onset of crystallization.

The Density Functional (DF) formalism, with advanced models for 
the non-local functional dependence of the Helmholtz free energy 
on the density distribution, has provided a
good framework to study the solid-liquid transition and other
highly structured systems. There have
been previous attempts to derive a Dynamic Density Functional (DDF) 
theory from the microscopic equations of motions \cite{Dean}-
\cite{Kirkpatrick2},
so that the density functional approximations developed for 
systems at equilibrium might be extended to the
dynamics of these systems. However, none
of these proposals is fully satisfactory as we shall demonstrate,
while the derivation that we consider
makes direct contact with the equilibrium DF formulation and is therefore 
consistent with thermodynamic requirements.

The theoretical foundations of the density functional
methods are based on the concept
that the intrinsic Helmholtz free energy of a fluid that exhibits
a spatially varying equilibrium density $\rho(r)$, is a unique
functional
$F[\rho]$ and is independent of the applied external fields for a given
intermolecular potential. An exact knowledge of $F[\rho]$ allows 
obtaining in a self-consistent fashion the  profile $\rho(r)$ and
all the n-point correlations via functional differentiation.
When dealing with
non-equilibrium situations, caused by
some changes of the external constraints,
such as the temperature, the pressure, or  electric field, 
it would be extremely useful to have similar methods
at our disposal.
Mode Coupling theories \cite{Goetze} provide
a kinetic approach to the dynamic of supercooled fluids
and structural glass transitions, but fail to predict the
crystallization process. In
in other cases the phase ordering dynamics
of liquids has been based
on schematic model Hamiltonians
of the Ginzburg-Landau type, which neglect the microscopic structure
or on heuristic approximations  for the free energy \cite{Wild}.

In principle the density is not the only relevant variable in a dynamical
description. The velocity distribution and correlation are crucial in the
understanding of hydrodynamic modes, while they do not appear in the 
equilibrium DF for classical fluids. However, one could argue following 
Cohen and de Schepper \cite{Cohen} 
that when the density is large, the momentum and the energy
flow quickly through the system via collisions, while the density variable
decays slowly. The hydrodynamic modes should become irrelevant for the
dynamics of dense and strongly structured systems; the only relevant variable
should be the density distribution, as in the equilibrium case, and 
the use of the equilibrium $F[\rho]$ is a promising starting
point to include the effects of the density correlations.

In the present study we restrict ourselves to systems with such
"relaxational dynamics", in which the velocity distribution
plays no relevant role. Instead of starting directly with the Newtonian
dynamics of the particles, we force the 
irrelevance of the velocity distribution at microscopic level
and begin from the stochastic Brownian equations of motion of 
a system of $N$ particles interacting via two-body forces.
In the equilibrium limit, as a very long time 
average of the dynamic evolution, 
the Newtonian and the Brownian equations of motion should 
give the same results and be equal to those of the 
equilibrium statistical ensemble. The relaxational dynamics
of dense systems has also to be similar for the two 
types of microscopic dynamics: the rapid flow of momentum and energy
due to particle-particle collisions in the Newtonian dynamics,
is given (most efficiently) by the bath in the Brownian dynamics. 
Of course, there are cases (including some of the examples
analyzed in this work) for which the Newtonian and the Brownian
dynamics have very different results. In those cases our
proposed DF approach to the dynamics would still be useful 
for systems following the Brownian equation of motion, like
colloidal particles in a bath, but it would not be appropriate
for systems in which the microscopic dynamics is Newtonian and
the velocity distribution becomes relevant.

The paper is organized as follows: in section 2 we derive the dynamic density 
functional theory starting from the stochastic equations of motion
of the particles and discuss the main features of the resulting DDF
approach. In section 3 we apply the method to few systems of
hard molecules in one dimension, for which the exact equilibrium free
energy density functional is known; the comparison between the DDF and 
the averages over the Langevin simulations gives a clear view of the
validity of our proposal. In section 4 we explore the same systems
but using now approximate forms for $F[\rho]$, of the same type as
those developed for realistic systems in three dimensions; in this
way we analyze the effects of the approximations used for $F[\rho]$,
which would be unavoidable in the practical use of the DDF formalism.
In 5 we draw the conclusions, discuss previous approaches and 
present the future perspectives.

\section{Transformation from
brownian trajectories to the equation for the density variable}

One considers an assembly of $N$ Brownian particles of coordinates ${\bf r_i}$
interacting 
via an arbitrary
pair potential $V({\bf r_i}-{\bf r_j})$ and experiencing
an external field $V_{ext}({\bf r_i})$ . Neglecting the inertial term and 
the hydrodynamic interaction their motion 
can be described by the following 
set of coupled stochastic equations:
\begin{equation}
\frac{d {\bf r}_i(t)}{d t}=- \Gamma {\bf \nabla_i } [\sum_j 
V({\bf r_i}-{\bf r_j}) + V_{ext}({\bf r_i})]
+{\bf\eta}_i(t)
\label{eq:uno}
\end{equation}
where the term ${\bf \eta}_i(t)= 
(\eta^x_i(t), \eta^y_i(t), \eta^z_i(t))$
represents the influence of the thermal 
bath and has the properties:
\begin{equation}
<\eta^{\alpha}_i(t)>=0
\label{eq:due}
\end{equation}
and
\begin{equation}
< \eta^{\alpha}_i(t)\eta^{\beta}_j(t') >=
 2 \ D \  \delta_{ij}\delta^{\alpha \beta}
\delta(t-t')
\label{eq:tre}
\end{equation}
where the average is over the  Gaussian noise distribution 
and $\alpha,\beta$ run over $x,y,z$.
The constants $\Gamma$ and $D$ give the mobility and the 
diffusion coefficient of the particles, respectively. The
Einstein relation gives $\Gamma/D= \beta \equiv 1/T$, and from
here on we take $\Gamma=1$ to fix the unit of time
and have $D= T = \beta^{-1}$. 
The evolution law drives the system towards the equilibrium situation
which is described by the canonical Gibbs probability measure. 
 Instead  of considering all the trajectories generated from eq.
(\ref{eq:uno}) we shall consider the evolution of the density of particles. 

 In order to render the paper self-contained we rederive briefly
the transformation \cite{Dean}, using the rules of the 
Ito stochastic calculus. In order to do so, we recall that 
if $f$ is  an arbitrary function of $x(t)$ given by the process:
\begin{equation}
\frac{d x}{d t}=a(x,t)+b(x,t) \xi(t)
\label{eq:gard1}
\end{equation}
with $<\xi(t)\xi(t')>=2 \delta(t-t')$
its evolution is given by the following Ito prescription
for the change of variables (see ref. \cite{Gardiner}):
\begin{equation}
\frac{d f(x,t)}{d t}=a(x,t) \frac{d f(x,t)}{d x}+b(x,t) 
\frac{d f(x,t)}{d x} \xi(t)+\frac{1}{2}(b(x,t))^2 
\frac{d^2 f(x,t)}{d x^2}
\label{eq:gard2}
\end{equation}
Thus employing eqs. (\ref{eq:gard2}) and (\ref{eq:uno}) we obtain:
\begin{equation}
\frac{d f({\bf r}_i(t))}{d t}=-{\bf \nabla_i } \left[\sum_j 
V({\bf r_i}-{\bf r_j})+ V_{ext}({\bf r_i}) \right]
\nabla_i f({\bf r_i})+\nabla^2_i f({\bf r_i}) 
+{\bf \nabla_i} f({\bf r_i}) {\bf\eta_i(t)}
\label{eq:cinque}
\end{equation}
 After inserting the identity $f({\bf r_i})=\int \delta({\bf r_i-r})
f({\bf r}) d {\bf r}$ and using the arbitrariness of $f$,
we obtain the equation for the partial density operator, 
$\hat\rho_i({\bf r},t)\equiv \delta({\bf r_i-r})$:
\begin{equation}
\frac{\partial \hat\rho_i({\bf r},t)}{\partial t}
=
T {\bf \nabla}^2 \hat\rho_i({\bf r},t) +
 {\bf \nabla} \left[
\hat\rho_i({\bf r},t) \left( \int d{\bf r'}   
(\hat\rho({\bf r'},t) {\bf \nabla} V({\bf r}-{\bf r'})+
{\bf \nabla} V_{ext}({\bf r}) \right) \right] 
+{\bf \nabla} \eta_i({\bf r},t) 
\hat\rho_i({\bf r},t)
\label{eq:sei}
\end{equation}
where, after some manipulation,
the instantaneous global density operator, $\hat\rho({\bf r},t)=
\sum_{i=1,N} \delta({\bf r_i(t)-r})$, can be shown to obey
the following multiplicative noise equation:
\begin{equation}
\frac{\partial \hat\rho({\bf r},t)}{\partial t}
=
{\bf \nabla} \left[
T {\bf \nabla} \hat\rho({\bf r},t) +
\hat\rho({\bf r},t){\bf \nabla}V_{ext}({\bf r})+
\hat\rho({\bf r},t) 
\int d{\bf r'}   
\hat\rho({\bf r'},t){\bf \nabla }V({\bf r}-{\bf r'}) 
+\eta({\bf r},t) 
\sqrt{\hat\rho({\bf r},t)} \right] .
\label{eq:sette}
\end{equation}
Eq. (\ref{eq:sei}) is a mere rewriting of the original eqs (\ref{eq:uno})
and as it stands does not represent an improvement. Also eq. 
(\ref{eq:sette}) is a stochastic equation for the total instantaneous density
$\hat\rho({\bf r},t)$ and is not an average, thus it is a sum of $N$ spikes
located at the positions  of the particles at the positions of the 
particles at time $t$. The connection with the DF formalism was made
by Dean \cite{Dean} in the following terms: the 
Helmholtz free energy density functional,
\begin{equation}
F[\rho]= T \int d {\bf r} \rho({\bf r}) [\log(\rho({\bf r})) -1]
+ \int d r \rho({\bf r}) \ V_{ext}({\bf r}) + \Delta F[\rho],
\label{eq:F}
\end{equation}
contains the exact ideal gas entropy and the external potential contribution
in the first two terms, while the third one includes the effects
of interactions and correlations between the particles and its exact
form is known only for very few systems.  The first two terms in the 
bracket of (\ref{eq:sette}) correspond precisely to the contributions
of the ideal gas and the external potential to $\rho \nabla \delta
F/\delta \rho$, and  
within an implicit mean field approximation it was
observed that the third term in the bracket of   
(\ref{eq:sette}) can also be cast in terms
of the functional derivative of $\Delta F[\rho]$.
In a slightly different language the analysis by  
Kawasaki \cite{Kawa1,Kawasaki} led to the
same type of proposal: a
dynamic density functional equation for $\hat{\rho}({\bf r}, t)$
in term of the functional derivative of $F[\rho]$ and a 
remaining stochastic contribution, from the noise term in (\ref{eq:sette}).
However, the use of the equilibrium functional $F[\rho]$
only makes sense with an ensemble averaged density distribution. 
The delta-function peaks in 
$\hat{\rho}({\bf r}, t)$
would give infinite contributions to the first term in (\ref{eq:F}),
because they correspond to a single microscopic state and not any
statistical ensemble average. To make the connection
between (\ref{eq:sette}) and a density functional description 
one has to implement some kind of averaging over the instantaneous
distribution of particles.  

 Within the microscopic Brownian dynamics
the obvious way to proceed is to average over the realizations of the
random noise $\eta({\bf r},t)$. We denote by brackets, $<....>$, the
results of this averaging and in particular we define the 
noise-averaged density $\rho({\bf r}, t) \equiv < \hat{\rho}({\bf r}, t)>$.
In the equilibrium limit, when the system has been allowed to relax for
long enough time under the Brownian dynamics, this average would 
correspond precisely to the Gibbsian equilibrium average. In the 
study of the dynamics, out of equilibrium, the density 
$\rho({\bf r}, t)$ 
has to be interpreted as an ensemble average, as in a collection
of colloidal systems with the colloidal particles at the same 
initial conditions but with different (thermalized) microscopic 
states for the bath. 
We may hope, without demonstration, that this Brownian ensemble
may also give an accurate description of dense systems with 
Newtonian dynamics near the crystallization, for which the 
ensemble should be interpreted as a collection of systems with 
the same initial positions but different (thermalized) velocities 
for the particles.

 When we proceed to take the noise average over the Brownian
evolution equation (\ref{eq:sette}) the first effect is to
cancel out the noise contribution in the last term. The 
stochastic equation for time evolution of $\hat\rho({\bf r}, t)$ 
becomes a deterministic equation for $\rho({\bf r}, t)$,
\begin{equation}
\frac{\partial \rho({\bf r},t)}{\partial t}
={\bf \nabla} \left[\int
T {\bf \nabla} \rho({\bf r},t) +
\rho({\bf r}, t) {\bf \nabla} V_{ext}({\bf r}) +
\int d{\bf r'}  <\hat\rho({\bf r},t) \hat\rho({\bf r'},t)>
 {\bf \nabla }  V({\bf r}-{\bf r'})\right],  
\label{eq:otto}
\end{equation}
in contrast with previous authors \cite{Dean},
\cite{Kawasaki}, \cite{Kirkpatrick},
who maintain the stochastic character of the dynamic DF evolution
keeping a random noise term \cite{footnote}. 

Systems of non-interacting ideal particles, 
$V({\bf r}-{\bf r'})=0$, provide 
an exact test of this point. In that case the free energy
density functional $F_{id}[\rho]$ reduces to the first two 
contributions in (\ref{eq:F}), since $\Delta F[\rho]=0$,
and (\ref{eq:otto}) may be written as a closed deterministic
equation for the density distribution,
\begin{equation}
\frac{\partial \rho({\bf r},t)}{\partial t}
= T \nabla^2 \rho({\bf r},t) + {\bf \nabla} \left[
\rho({\bf r},t){\bf \nabla} V_{ext}({\bf r}) \right]
={\bf \nabla} \cdot \left[
\rho({\bf r},t){\bf \nabla } \frac{\delta F_{id}[
 \rho({\bf r},t) ]}{\delta \rho({\bf r},t)} \right], 
\label{eq:DFid}
\end{equation}
which is the exact Fokker-Planck equation for the diffusion
and drift of an ideal Brownian gas. In the proposals of
Dean \cite{Dean} and Kawasaki \cite{Kawasaki}
this equation would include a stochastic noise 
term, leading to an overcounting of the
fluctuations. In particular the Boltzmann equilibrium 
state predicted by (\ref{eq:DFid}) in the static limit
would be spoiled by the presence of the random noise, 
which would be equivalent to an overestimation of the 
temperature.

 In the case of interacting particles equation
(\ref{eq:otto}) is not a closed relation, since in order to obtain 
$\rho({\bf r},t)$ one needs the equal-time two-point
correlation $\rho^{(2)}({\bf r},{\bf r'},t) \equiv 
<\hat\rho({\bf r},t) \hat\rho({\bf r'},t)>$. 
The simplest mean field approximation 
assumes $\rho^{(2)}({\bf r},{\bf r'},t) \approx \rho({\bf r},t) 
\rho({\bf r'},t)$ and gives a closed equation for    
$\rho({\bf r},t)$, but it would give quite pathological
results for the molecular core repulsions. 
Following the same procedure used for $\rho({\bf r},t)$
 we may obtain an equation
for the time evolution of $\rho^{(2)}({\bf r},{\bf r'},t)$
which in turn depends on 
the three-point correlation. In fact eq. (\ref{eq:otto})
is only the first member of an infinite hierarchy of relations
known as Born-Bogolubov-Green-Kirkwood-Yvon (BBGKY) 
integro-differential equations
connecting n-point functions to (n+1)-point functions
\cite{Hansen}. As in the 
equilibrium case one can get approximated results breaking the chain
at any level. The so called Kirkwood superposition approximation, 
replaces the three-point correlation by a product 
of two-point correlations as the next step after
the mean field approximation and may already give reasonable results
for hard core interactions.
 
Here we propose a different strategy, a density
functional approach in which the two-point correlation function may
be approximated with the help of equilibrium free energy density 
functionals. The excess free energy density functional $\Delta F[\rho]$
in principle contains all the equilibrium
correlation structures in the system and, although the exact functional form
is known only for very few systems, there are workable and very 
accurate approximations for most systems of interest. We may use
the information contained in $\Delta F[\rho]$, about the 
correlation structure at equilibrium, to approximate      
$<\hat \rho({\bf r},t) \hat\rho({\bf r'},t)>$ in a system out 
of equilibrium. In this way we get a generic, closed, Dynamic Density 
Functional relation for $\rho({\bf r},t)$ equivalent to (\ref{eq:DFid})
with a clear interpretation of the approximations involved.

Let us consider an equilibrium state of the system characterized 
by an arbitrary profile $\rho_0({\bf r})$ (the subscript $0$ indicates
the equilibrium average),
which we shall eventually take equal to the profile $\rho({\bf r},t)$
at a given instant $t$. Such an equilibrium state certainly exists and 
represents a minimum of the grand potential functional provided we add
an appropriate equilibrating external potential. One can prove, in fact,
that  for fixed temperature, chemical potential, and 
pair interactions there always exists a unique external potential 
$u({\bf r})$ which induces the given $\rho_0({\bf r})$. 
In other words, upon adding the external potential
$u({\bf r})$, we would pin the system to be at equilibrium in a 
configuration corresponding to the instantaneous average density 
$\rho({\bf r},t)$; the potential $u({\bf r})$ 
is a functional of $\rho_0({\bf r})$ and changes with $t$ as 
$\rho({\bf r},t)$ varies \cite{Evans}.  

 From the general properties of the equilibrium functionals we have that 
the following two exact equilibrium relations must be satisfied by 
$\rho_0({\bf r})$ and $u({\bf r})$. First, the local balance of
momentum at any point implies the BBGKY relation, 
\begin{equation} 
\frac{1}{\rho_0({\bf r})}{\nabla \rho_0({\bf r})
+\beta \nabla[V_{ext}({\bf r})+ u(\bf r})]
=-\beta \frac{1}{\rho_0({\bf r})}\int d r' 
\rho^{(2)}_0({\bf r, r'}) \nabla V({\bf r-r'}).
\label{eq:BBGKY}
\end{equation}
 Second, the thermodynamic equilibrium implies that the
functional derivative of $F[\rho]$ at any point is equal to a
uniform chemical potential $\mu$. Taking the gradient gives the
equation, first obtained by Lovett et al. \cite{Evans}, 
\cite{Lovett},

\begin{equation}
\frac{1}{\rho_0({\bf r})}{\nabla \rho_0({\bf r})
+\beta \nabla [V_{ext}({\bf r})+u(\bf r})]=
-{\bf \nabla}\frac{\delta}{\delta \rho_0({\bf r})}[\beta 
\Delta F[\rho_0]] =\int d {\bf r'} 
c^{(2)}({\bf r},{\bf r'}) \nabla \rho_0({\bf r'}),
\label{eq:lovett}
\end{equation}
where
$$
c^{(2)}({\bf r},{\bf r'}) = -\beta
\frac{\delta^2 \Delta  F[\rho_0] } {
 \delta \rho_0({\bf r}) \delta \rho_0({\bf r'})} 
$$ 
is the direct correlation function, related to the 
functional inverse of the equilibrium two point 
density-density correlation $\rho^{(2)}_0({\bf r, r'})$.  
The functional $\Delta F[\rho_0]$ serves to generate the sequence
of inverse linear response or direct correlation functions \cite{Lebowitz}
upon functional differentiation with respect to $\rho$.
Let us emphasize that eqs (\ref{eq:BBGKY}) and (\ref{eq:lovett}) are exact 
for the instantaneous potential $u({\bf r})$. 
Comparing these two equations and assuming that the 
equal time correlation $\rho^{(2)}({\bf r,r'},t)$, averaged over 
the Brownian noise, may be approximated by that of the equilibrium system
with the same density distribution, we get the last term in
eq. (\ref{eq:otto}) as:
\begin{equation}
\int d{\bf r'}  <\hat\rho({\bf r},t) \hat\rho({\bf r'},t)>
 {\bf \nabla }  V({\bf r}-{\bf r'}) =  
\rho({\bf r},t){\bf \nabla } \frac{\delta \Delta F[
 \rho({\bf r},t) ]}{\delta \rho({\bf r},t)} .
\label{eq:DF1}
\end{equation}

In summary, we have used the fact that at any instant we can find 
a fictitious
external potential $u({\bf r})$ which  equilibrates 
the system, i.e. constrains
its grand potential to be  minimal. This minimum is characterized by
the imposed density profile $\rho_0({\bf r})=\rho({\bf r},t)$ 
and by equilibrium correlations $\rho_0^{(2)}({\bf r,r'})$ consistent 
with it.  The present approximation
replaces the true off-equilibrium pair distribution function
$<\hat\rho({\bf r},t) \hat\rho({\bf r'},t)>$ by the equilibrium
$\rho_0^{(2)}({\bf r,r'})$, and then uses the equilibrium density
functional $\Delta F[\rho]$ to obtain the relevant information on
this function.

The assumption that the two routes, 
eqs. (\ref{eq:BBGKY}) and (\ref{eq:lovett}),
are equivalent implies that the 
Fluctuation Dissipation theorem holds, while in general, 
out of equilibrium,
it is violated \cite{Kurchan}. In fact the relation connecting 
$\rho_0^{(2)}({\bf r,r'})$ 
to $c^{(2)}({\bf r,r'})$  (the O.Z equation) is an exact equilibrium 
property and
is based on the idea that the correlation function is the matrix inverse
of the second derivative of the functional $F$ with respect to 
$\rho_0({\bf r})$.

With (\ref{eq:F}) and (\ref{eq:DF1}) we may recast (\ref{eq:otto})
into the main result of the Dynamic Density Functional approach, 
based on the use of the equilibrium functional $F[\rho]$:
\begin{equation}
\frac{\partial \rho({\bf r},t)}{\partial t}
={\bf \nabla} \cdot \left[
\rho({\bf r},t){\bf \nabla } \frac{\delta F[
 \rho({\bf r},t) ]}{\delta \rho({\bf r},t)} \right], 
\label{eq:DF}
\end{equation}
which has the form of a continuity equation, $\partial \rho/\partial t+
\nabla \cdot {\bf j}=0$, with the current of particles given by:
\begin{equation}
{\bf j}(r,t)= - \rho(r,t) {\bf \nabla} 
\left. \frac{\delta F[\rho]}{\delta \rho(r)}\right|_{\rho({\bf r},t)}. 
\label{eq:current}
\end{equation}

The main features of this approach are the following:

a) $F[\rho]$ is a functional solely of the density field 
and thus eq. (\ref{eq:DF}) is a closed non linear 
equation for $\rho(\bf{r},t)$.

b) The equation is deterministic, but the variable
$\rho(\bf{r},t)$ has to be interpreted as the
instantaneous density operator averaged over the 
realizations of the random noise $\eta_i(t)$.
The contribution from the ideal gas  entropy generates the
diffusion term in (\ref{eq:DFid}) and reflects the
presence of the thermal noise.

c) The only assumptions leading to (\ref{eq:DF}) are that
the systems follows a relaxative dynamics, which may
be described by the Brownian motion of the particles 
in a thermalized bath, and that the instantaneous
two-particle correlations are approximated by those
in an equilibrium system with the same density 
distribution, as given by the (exact or approximated)
density functional $F[\rho]$.

Before concluding this section we note an interesting feature of the
dynamics: using the equation of evolution 
for $\rho({\bf r},t)$ it is straightforward to show
that for any system with closed or periodic boundary conditions, 
the dynamics always tends to decrease the free energy
functional, i.e.
\begin{equation}
\frac{d F[\rho]}{dt}=-\int d {\bf r} \ \rho({\bf r},t) \ \left[
\nabla \frac{\delta F}{\delta \rho} \right]^2 
\leq 0.
\label{eq:decrease}
\end{equation}
In the long time limit, the evolution of the system leads
to its equilibrium density distribution, which
corresponds to a uniform value of
$\mu=\delta F/\delta \rho ({\bf r})$, i.e. the usual 
Euler-Lagrange equation in the equilibrium DF formalism. 
However, the trajectories that lead to the minima of $F$ are not 
necessarily along the directions corresponding to the maximum slope. 
The continuity equation, implies that the local conservation 
of particles is built in and imposes important constrains 
on the local changes of $\rho(\bf{r},t)$.

For any system with a finite number of particles there is a unique
canonical equilibrium density distribution, $\rho_0({\bf r})$, which
corresponds to the unique local (and global) minimum of the exact 
free energy density functional $F[\rho]$. However, the use of approximations
for $F[\rho]$ may lead to the existence of several local minima
in which the dynamics of eqs. (\ref{eq:DF},\ref{eq:decrease}) may get
trapped; this deserves further comment. Notice first that these 
local minima of the free energy density functional, and the barriers 
between them,
cannot be directly associated to the local minima, and 
to the barriers,
of the potential energy in the Langevin description. Consider 
the dynamics of an ideal Brownian gas in an external potential 
with two local minima,
separated by a barrier $V_b$; the Langevin representation requires
the gaussian noise $\eta_i(t)$ to allow the particles to go over the
barrier. The probability of such a jump is proportional to $\exp(-V_{b}/T)$ 
and, for large barriers, it sets the scale of time for the equilibration 
of the system. The same time scale appears in the exact Fokker-Planck 
representation (\ref{eq:DFid}) through a different mechanism:
the free energy landscape has a single minimum, at the equilibrium
density $\rho_0({\bf r}) \sim  \exp(-V_{ext}({\bf r})/T)$, there
are no free-energy barriers and the deterministic time evolution 
is set by the particle current (\ref{eq:current})
along the functional path to equilibrium. 
When the equilibration requires moving particles
across a large potential barrier, the relaxation time is
imposed by the density there,
$\rho({\bf r},t) \sim exp(-V_b/T)$, which may produce a
very weak current even if the gradient of the local chemical 
potential is large.

   The presence of different local minima in approximate free
energy density functionals is often found in the use of 
the density functional formalism for equilibrium properties.
The usual interpretation is that the global 
minimum gives the true equilibrium state, while other
local minima are associated with metastable states, 
and phase transitions
are described as the crossover of different minima
\cite{Wolynes}. This 
is a (generalized) mean-field level of description, in which
the phase transition is described in terms of an order parameter
related to the one-particle distribution functions. The exact 
description should always give
convex thermodynamic potentials and it requires
the description of the phase transitions in terms of the 
N-particle correlations, rather than in terms of a one-particle
order parameter. 
  
 In the present context, the use
of approximate free energy density functionals with more than 
one local minima poses a problem of interpretation. When the 
deterministic time evolution (\ref{eq:DF})  gets trapped  at
a 'metastable state' it can never reach the true
equilibrium state.  In previous attempts to use $F[\rho]$
for dynamics, both Dean \cite{Dean} and Kawasaki \cite{Kawasaki}
have kept an extra random noise term in (\ref{eq:DF}), as a
remnant of the original random noise in the Langevin dynamics.
This random noise, allows the system to jump over any 
'metastability barrier' in a finite time but it leaves the
problem of identifying its origin and intensity. From our
analysis it is clear that the use of the free energy density functional
makes sense only for the density $\rho({\bf r},t)=<\hat\rho({\bf
r},t)>$, averaged over the realizations of $\eta_i(t)$ in the 
Langevin dynamics, and this averaging gives the deterministic
equation (\ref{eq:DF}). We proceed here to explore the 
results of the deterministic DDF formalism and come back to this 
point in the conclusion section.

The practical use of (\ref{eq:DF}) to study problems like the 
growth of a liquid drop from an oversaturated vapour 
seems to be limited. The classical Lifshitz-Slyozov-Wagner theory
\cite{LSW} for the late stages of 
growth, beyond the critical droplet size, is given directly from
(\ref{eq:DF}), even if we use the simplest local density approximation
for $F[\rho]$. However, any
available approximations for $F[\rho]$, 
will be unable to describe the early stages of nucleation, as 
they do not include the effects of long-range critical-like correlations.
However, we now have density functional approximations with a
good description of the short range correlation structure in
highly packed systems. The DF description of the freezing of a
liquid \cite{WDA},\cite{Denton}
or the study of fluids confined to narrow pores \cite{Us} are
the most remarkable achievements of the non-local 
density functionals developed in the last decades. We believe 
that the DDF formalism may be used to study the dynamics of densely
packed fluids taking full advantage of the description of the
correlation structure at short range, given by suitable approximations
for the equilibrium $F[\rho]$. The most interesting (and difficult)
problem would be the study of a closely-packed system, which has
not crystallized. The Langevin description (\ref{eq:uno}) may
give extremely slow dynamics since either the random noise
has to take the system over large energy barriers (like in the 
ideal gas described above), or it has to produce rearrangements 
which are only possible through unlikely correlations of many 
particles. Within the deterministic dynamic density functional 
approach (\ref{eq:DF}) this situations may be seen in two possible
ways:

1) The free energy landscape becomes rough and displays many local
   minima, whose number grows exponentially with the size of the system
  \cite{Parisi},
   and within our description the system would remain indefinitely 
   trapped in any of these minima, unless it is annealed at a 
   higher temperature.
   The idea is that when a uniform liquid 
   is forced to have a density larger than the one corresponding to a liquid
   at coexistence with the solid, the system starts developing inhomogeneous
   patterns, which are associated with local minima of the free energy.   
   One conjecture relates the origin of the glassy 
   behavior in liquids to the existence of these minima, as an extension
   of the well founded theory of mean-field spin-glasses \cite{Kirkpatrick},
  \cite{Parisi}. 
 
2) Alternatively, the free energy landscape $F[\rho]$ may be smooth, 
   with a single minimum, because the correlations 
   required to relax the system 
   are described well enough by the free energy functional.
   In this case, the system will never get really frozen, until it reaches 
   the
   true equilibrium state, but the dynamics may become 
   so slow that it may appear to be frozen in any practical computation.
   The slow dynamics of (\ref{eq:DF}) may be a result of having very low
   density along current path (again, as in the ideal gas above)
   or of requiring very unlikely (but not impossible) correlations 
   of many particles.  This case would correspond to the conjecture that 
   the glassy behaviour in liquids is due to a divergence in the 
   viscosity, and in the relaxation times, rather than to the
   existence of a 'metastable freezing'.

Within the dynamical density functional approach having one or the
other way would be a result of using a worse of a better approximation
for $F[\rho]$. The configurations with a very large escape time, when
described by a good density functional, may appear like permanent
stable states when the dynamics is described with a poorer
approximation for $F[\rho]$. In the next section, we present 
explicit results for a simple model in one dimension which 
displays these features and illustrates how the structure of
the nearly-frozen states may be given by simple DF approximations,
while the actual calculation of the escape time may require a
density functional giving good account of the many-particle
correlations.

\section{Applications to 1-d problems employing the Exact functional}

Since neither the correctness nor
 the feasibility of the approach presented  have  been 
tested so far, we shall begin comparing the two levels of description: the 
Brownian dynamics and DF dynamics and in order not to introduce
unnecessary sources of discrepancies we consider a one dimensional
hard-rod system , whose exact equilibrium  DF is known \cite{Percus},
\cite{Robledo}
and which includes strong correlation effects due to the infinite
repulsion between the particles.

Long ago, Percus \cite{Percus} was able to determine the exact form
of the  free energy density functional for an assembly of hard rods
of length $\sigma$.
The excess functional due to the interactions reads:
\begin{equation}
\Delta F[\rho(x)] =  
\int_{-\infty}^{\infty} dx \ \phi(\eta(x)) \
\frac{\rho(x+\sigma/2)+\rho(x-\sigma/2)}{2},
\label{eq:percus}
\end{equation}
where
\begin{equation}
\phi(\eta)=- T \ log(1-\eta(x)),
\label{eq:nonideal}
\end{equation}
and the local packing fraction $\eta$ is defined as: 
\begin{equation}
\eta(x)=\int_{x-\sigma/2}^{x+\sigma/2} dx' \rho(x').
\label{eq:packing}
\end{equation}
There are other equivalent
ways to calculate $F$ (like taking the weighting graining only towards
the right, or towards the left) which yield the same value of $F$
for any density distribution. Using eq. (\ref{eq:DF}) 
we obtain the following equation for the time evolution of the density:
\begin{equation}
\frac{\partial \rho(x,t)}{\partial t} = 
\frac{\partial^2 \rho(x,t)}{\partial x^2}
                 +\frac{\partial}{\partial x} 
\left[\rho(x,t)\left(\frac{\rho(x+\sigma,t)}
                 {1-\eta(x+\sigma/2,t)}
           -\frac{\rho(x-\sigma,t)}{1-\eta(x-\sigma/2,t)}
\right)\right]
                  + \frac{\partial }{\partial x} 
 \left[ \rho(x,t) \frac{d V_{ext}(x)}{dx} \right],
\label{eq:rho}
\end{equation}
where we have chosen the energy units such that $T=1$.

The first term in (\ref{eq:rho}) represents the diffusion equation for
ideal gas case, the second term is the correction due to the hard-rod
interaction. It is worthwhile to point out that $\rho(x)\rho(x+\sigma)/
(1-\eta(x+\sigma/2))$ is just the two-point equilibrium correlation
function $\rho_0^{(2)}(x,x')$ evaluated at contact, i.e. when
$x'=x+\sigma$, so that this term takes into account the collisions of the
rod at $x$ with the remaining particles on the right hand side. Similarly
the other term describes the interactions with the left sector.

When the density
profile varies very smoothly (compared with the hard-rod length)
the second term may be written in terms of the local chemical
potential and the compressibility and one obtains a diffusion equation
with a renormalized constant.

\subsection{Free expansion from a dense state}

\begin{figure}
\centerline{
\psfig{figure=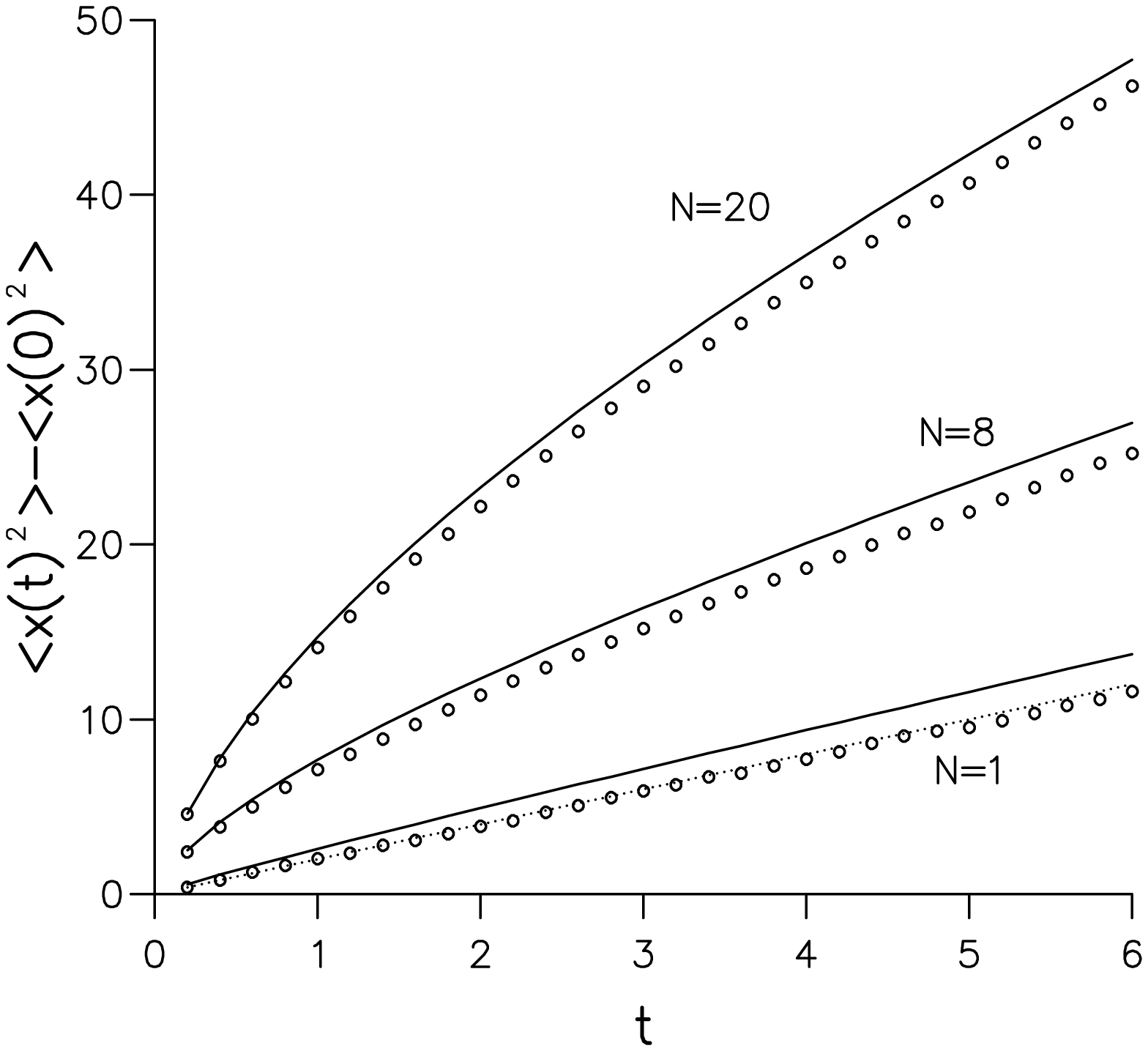,width=10.cm,height=10.cm,angle=-90}}
\caption{Density profiles for a system of $N=8$ hard rods of
unit length in free expansion, at different times. The full
lines are the results of the DDF equation and the dots are
the average over 2000 Langevin simulations. The results have
been shifted in the vertical direction to allow a clear view.
}
\label{fig:dos}
\end{figure}

Our first check has been to compare the results of 
Langevin simulations eq. (\ref{eq:uno})
with the results obtained by means of the DDF eq. (\ref{eq:DF})
and (\ref{eq:rho}),
using  the exact $F[\rho]$, for the free expansion of
$N$ hard rods in absence of external potential.
In Figure (\ref{fig:dos}) we present the density profiles
for a system of $N=8$ hard rods of unit length, $\sigma=1$.
The temperature is fixed to $T=1$ and in the initial configuration,
at $t=0$, the rods 
are set at fixed positions separated by a distance $1.05$ 
between their centers. In the first stages of the time evolution
each rod develops a gaussian density distribution, and the superposition
of all the rods gives a total density distribution with strong
oscillations, typical of tightly packed hard molecules. With
increasing time the packet expands, the oscillations become weaker
and then disappear. For very large $t$ (not shown in the figure)
the packet becomes very wide, with $\rho(x,t)<<1$ everywhere, so that
the collision term in (\ref{eq:rho}) is nearly irrelevant. In that
limit the system evolves like an ideal gas with a gaussian 
distribution of width proportional to $\sqrt{2 t}$.

\begin{figure}
\centerline{
\psfig{figure=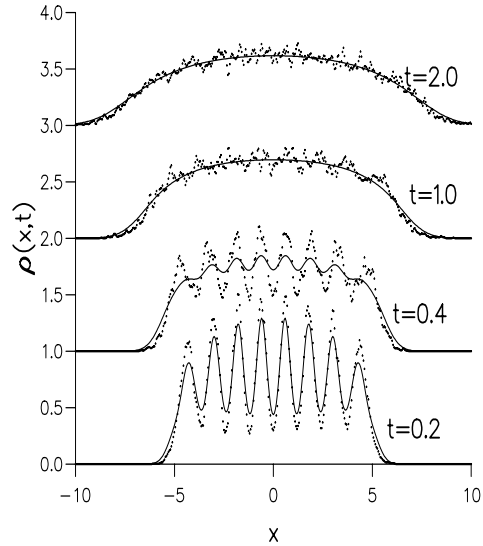,width=10.cm,height=10.cm,angle=-90}}
\caption{Mean squared displacements for systems of $N$ hard rods
of unit length, in free expansion from highly dense initial state.
The full lines are the results of the DDF equation and the open
circles are the average over 2000 Langevin simulations, for the 
respective values of $N=1$, $8$ and $20$. The dotted line
is the result for any number of ideal, non-interacting, particles.}
\label{fig:uno}
\end{figure}

The dots in Figure(\ref{fig:dos}) are the average over 2000 
Langevin simulations for the same system. The qualitative trend
is similar to the DDF results, but the 
damping of the oscillations is clearly slower in the simulation.
In Figure(\ref{fig:uno}) we present the time evolution of 
$<x(t)^2 - x(0)^2>$ for packets with $N=1$, $8$ and $20$,
to give a measure of the rate of expansion. The dotted line is the exact 
result for the ideal gas,  $<x(t)^2 - x(0)^2>= 2 \ t$, which is 
independent of $N$. For hard rods there is a clear enhancement 
of the effective expansion rate, because the rods at the two ends
of the packet have a strong bias towards moving away from the
the packet. The result of this effect increases with $N$ because
it acts until the whole packet has expanded. For $t>>1$ and
any value of $N$ the slope of $<x(t)^2 - x(0)^2>= 2 \ t$ 
goes to the ideal value, 
but the enhanced expansion at small $t$ produces a shift of the 
values with respect to the ideal gas. The comparison between the DDF 
and the average over 2000 Langevin simulations again shows the
same overall trend and also the same dependence with $N$, but
the DDF gives always a slightly larger expansion rate.
The case with $N=1$ offers a clear explanation for this discrepancy:
with a single rod in the system there should be no collisions 
and the results should be those of the ideal gas, but the second term
in the DDF equation (\ref{eq:rho}), obtained with the exact $F[\rho]$,
still gives a contribution, unless $\rho(x,t) \rho(x+\sigma,t)=0$
for any $x$. The reason is that in the exact $F[\rho]$ (\ref{eq:percus}),
as for any density functional used in the DDF,  the density distribution
has to be interpreted in the grand-canonical ensemble. It corresponds
to a system in contact with a particle reservoir, in which the chemical
potential is set to give the average value of $N$; however, the configurations
contributing to the density distribution may have any number of 
particles. In the system with $<N>=1$, there would be contributions 
from the density distributions with
$N=2$, $3$,... (compensated with the contribution with $N=0$), and
these contributions include the effects of the collisions.

  In general, the fluctuation in the number of particles opens a 
relaxation path which is not present in the Langevin simulations,
carried with fixed $N$. This extra relaxation path produces the 
faster damping of the oscillations in the density profiles and the 
larger diffusion rate of the DDF. The effect is important
only in the intermediate stage of the expansion because for $t<<1$
the compressibility of the the system is too low to have important
fluctuations in $N$ and for $t>>1$ the system is so diluted that 
the total effect of collisions is negligible. 

\subsection{Collapse to a dense equilibrium state}

\begin{figure}
\centerline{
\psfig{figure=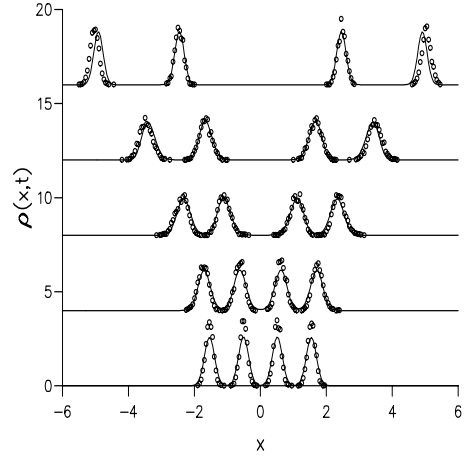,width=10.cm,height=10.cm,angle=-90}}
\caption{Density profiles of four hard rods of unit length, 
collapsing to the equilibrium state, in a parabolic potential. 
The full lines are the results of the DDF equation and the circles
are the average over 2000 Langevin simulations. The results have
been shifted in the vertical direction and they correspond, 
from top to bottom from $t=0.01$ to $t=0.09$, at $0.02$ intervals. 
The equilibrium state is very close to the later time results.
}
\label{fig:tres}
\end{figure}

In Figure(\ref{fig:tres}) we display the density profiles for 
a system of four hard rods falling to the bottom of a 
parabolic potential well, $\beta V_{ext}(x)=a x^2$, with $a=10$.
At $t=0$ the rods are located at $x= \pm 3$ and $x=\pm 6$, well
separated from each other. During the first stages of the time
evolution each rod follows a steady drift due to the external force
and, at the same time, they develop gaussian peaks of increasing 
width, due to the random diffusion. The collisions between the
two rods at each side of the potential well become important
for $t \approx 0.05$; later the two packets collide and 
relax to the equilibrium density distribution, which
is reached for $t=0.1$, within our numerical precision.
The results of the DDF and the average over 2000 Langevin
simulations are in good agreement, although small discrepancies
may be observed both in the early drifting peaks and in the 
final (equilibrium) profiles. The origin of these discrepancies
is again the difference between the Langevin simulation with
fixed $N$ and the grand-canonical DDF. The final equilibrium 
density distribution with the Langevin simulation corresponds
to the canonical ensemble and for systems with small number 
of particles it is known to be different from the grand-canonical
distribution \cite{Salamanca}. In our case, the difference 
depends on the value of the parameter $a$, in the external potential:
for $a>>1$ the rods are very tightly packed, the compressibility
is very low and the fluctuations in the
number $N$ are very small; for
$a<<1$ the final equilibrium profile is very broad, without 
oscillations, and the effect of the collision is too weak to
produce observable differences; but for intermediate values 
$a \approx 1$ the difference between the canonical and the
grand canonical ensemble may be quite important and it is
reflected in the different time evolution predicted by the DDF and 
by the Langevin simulations. 
 The use of a canonical $F[\rho]$
in the DDF equation (\ref{eq:DF}) would, in all probability, give a
better agreement between the two methods, but unfortunately
we are not aware of any explicit canonical density functional 
for interacting particles. Nevertheless, the use of the 
equilibrium $F[\rho]$ to include the role of the molecular
correlations is always an approximation, so that even without
considering the difference in the thermodynamic ensemble,
it was not obvious that the agreement between the DDF and the 
Langevin simulation would be as good as observed in the preceeding
figures.  

\subsection{ Relaxation through highly correlated states}

The third check of the theory, being considered, is the relaxation in a 
system which requires strongly correlated motions of all
the particles. We place the hard rods in a periodic external 
potential $V_{ext}(x)=-V_o \cos(2 \pi \ x)$, with minima
at any integer values of $x$. The  hard-rods length is taken
as $\sigma=1.6$, so that two rods cannot be at the
bottom of nearest-neighbour wells. We take periodic boundary conditions 
with total length $L=8$ and set $N=4$ rods, which 
at the initial time  are at the bottom of every second well of 
$V_{ext}(x)$. The equilibrium density distribution, which may be
obtained directly by the minimization of the exact free energy
density functional, has the full symmetry of the the external
potential, so that the relaxation process has to shift
(on average) half a particle from the initially occupied wells to
those wells which are initially empty. However, the jump of a 
rod over the barrier, to the next potential well, is not 
compatible with keeping the next rod at the bottom of its potential
well. The system has to pay the extra energy of keeping the two
consecutive rods away from the minima or it has to relay on a
correlated motion of the $N$ rods, to shift from one subset of 
minima to the other one.    
      
\begin{figure}
\centerline{
\psfig{figure=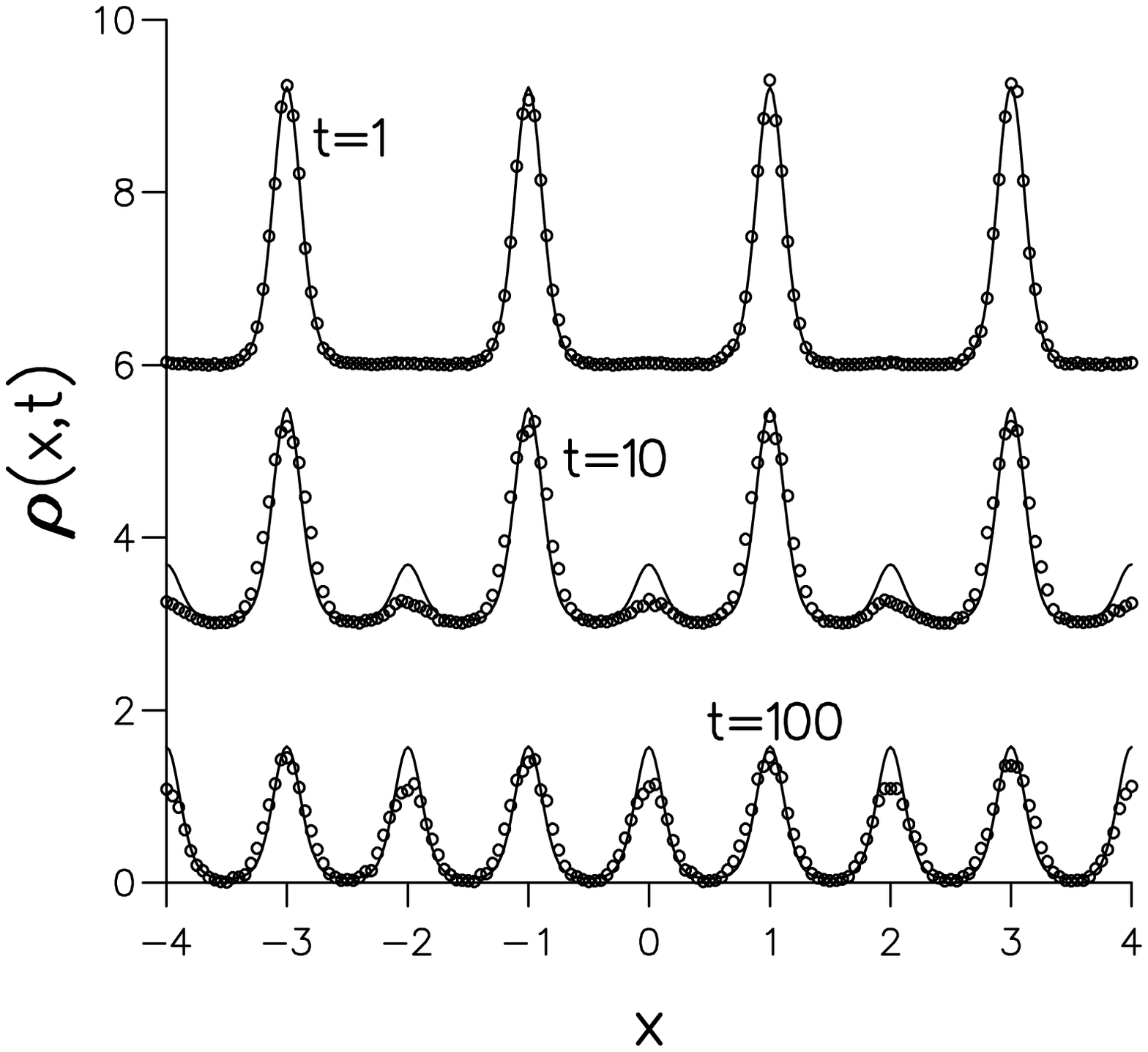,width=10.cm,height=10.cm,angle=-90}}
\caption{
Density profiles of four hard rods of length $\sigma=1.6$ in 
a periodic external potential with 8 minima separated by the
unit length. In the initial state the rods are in alternate 
minima, located at the odd integer values of $x$, and the system 
relaxes towards the equilibrium state in which all the 
potential wells are equally populated. 
The full lines are the results of the DDF equation and the circles
are the average over 2000 Langevin simulations. The results have
been shifted in the vertical direction.
}
\label{fig:cuatro}
\end{figure}

 In Figure (\ref{fig:cuatro}) we present the time evolution of
the density profiles with $\beta V_o=2.$ showing different times,
for both the DDF equation (\ref{eq:DF},\ref{eq:rho}) and the 
average of the Langevin simulations over $2000$ realizations
of the noise. In both cases the relaxation is slow (compared
with the previous examples) and it becomes much slower for increasing 
values of $\beta V_o$ or $\sigma$. In agreement with our
general prediction eq.(\ref{eq:decrease}), the system flows to the
unique equilibrium state, density peaks grow at the 
positions of the potential wells which were initially empty
until the exact equilibrium density profile is obtained.
The comparison of the result clearly shows that
the DDF equation, with the exact Percus free energy,
approaches the equilibrium
state faster than the average of the Langevin simulation.
This difference is related again to the use of different
statistical ensembles: the canonical Langevin equation
keeps constant $N$ while the grand-canonical $F[\rho]$
allows for fluctuations in the number of particles,
keeping only the average.  The changes in $N$ 
in the DDF open a new relaxation path and gives a faster
relaxation, even is the final equilibrium density
profiles in the canonical and the grand-canonical
ensembles are very similar.

\begin{figure}
\centerline{
\psfig{figure=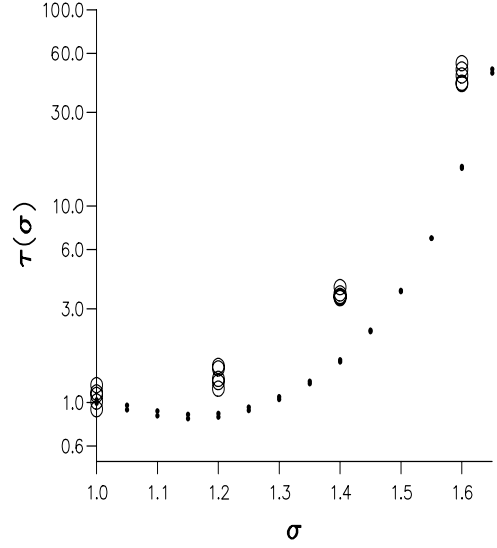,width=10.cm,height=10.cm,angle=-90}}
\caption{Relaxation time $\tau$ for the system in Figure (4) as
a function of the rod length $\sigma$, for fixed amplitude of the external
potential $\beta V_o=2.$, calculated
from the results of the DDF equation (dots) and the average over 2000 Langevin 
simulations (circles), the dispersion of the data reflects the results of 
$\tau=-t/log(\xi(t))$ at different times.
}
\label{fig:cinco}
\end{figure}

 To get a quantitative description of the relaxation times 
we define an order parameter which gives, at any time,
the relative difference
between the occupation of the odd and the even potential wells,
$\xi=(N_{odd}-N_{even})/N$. The initial condition
sets $\xi=1$ and the final equilibrium state corresponds
to $\xi=0$. Both in the case of the DDF and of the Langevin
dynamics we observe a pure exponential decay, $\xi(t)=exp(-t/\tau)$
so that whole process may be described by the relaxation time
$\tau$.  The importance of the particle correlations in
the slowing of the relaxation dynamics is shown in
Figure (\ref{fig:cinco}) through the
dependence of the relaxation time with on the size of the 
hard-cores, $\tau(\sigma)$, keeping the same external potential.
Both the Langevin dynamics and the DDF show a fast decrease of $\tau$,
by nearly two orders of magnitude, when the rod size decreases 
from $\sigma=1.6$ to $\sigma=1.$, which makes
the occupancy of neighbour potential wells easier.

The increase of the external potential amplitude, for fixed $\sigma=1.6$,
produces similar results, as presented in Figure
(\ref{fig:seis}). The relaxation time grows 
faster than exponentially with $\beta V_o$, and 
the results of the DDF follow from below the general 
trend of the Langevin dynamics, 
over several order of magnitude for $\tau$. For comparison we
present in the same figure the relaxation times for the ideal
gas in the same external potential.

 The conclusion is that the DDF formalism, with the
the exact equilibrium free energy density functional,
is able to reproduce qualitatively the low
relaxational dynamics of a system produced by the
packing constraints of molecular hard cores. In our example the
final equilibrium state, with equal occupancy of all the
potential wells, represents a superposition of
highly correlated states, in which all the particles 
are in alternate wells, either in the odd or the even positions. 
The relaxation from the asymmetric initial state requires
strong correlations, with all the particles moving 
together or in cascade, and the unlikelyness of these
correlations gives the long relaxation time.  
This process may be regarded as a (very simplified) simile of  
the relaxation from a very dense, non-crystalline
initial state: the approach to equilibrium may require the
correlated rearrangement of many molecules, the unlikelyness of
these correlations produces relaxation times which are so long that 
the system may appear as frozen in an ill crystallized state.

\section{Results with approximate density functionals}

The one-dimensional hard-rod system provides an interesting
test to the DDF, because we may use the exact equilibrium 
density functional.
Unfortunately, given the Hamiltonian describing a set of $N$ 
interacting particles in a d-dimensional space, 
the exact free energy functional is in general unknown, 
since the exactness would be equivalent to calculating the exact partition
function of the model in any external potential. Consequently one must
rely on suitable approximation schemes and, in the case of 
repulsive hard-core interactions, there exist several approximate
methods which yield accurate results \cite{Evans}.
The requirements imposed are: first, one must recover
the thermodynamic properties of the homogeneous fluid; second, one must
reproduce the structure of highly inhomogeneous systems;
finally, the approximation must  satisfy a number of exact 
relations.

Among the most successful density functionals approximations
for the free energy of hard spheres we have those known under the
generic name of weighted density approximations (WDA). The main assumption 
is that for each particle, at some point ${\bf r}$, there is a contribution
to the excess free energy, $\phi(\overline \rho({\bf r}))$,
which is a local function of a weighted density 
${\overline \rho}({\bf r})$, obtained by averaging the
true density profile $\rho({\bf r})$ over a small region centered at 
${\bf r}$:
\begin{equation}
{\overline \rho}({\bf r})=\int d{\bf r'} w(|{\bf r-r'}|,
{\overline \rho}({\bf r})) \ \rho({\bf r'}).
\label{eq:peso}
\end{equation}

 The excess free energy per particle, $\phi(\rho)$, is
obtained from the bulk equation of state and the weight function
is set to get a good (approximate) description of the bulk
correlation structure. Different recipes have been used and in 
general they require a density 
dependent weight function \cite{WDA},
\cite{Denton}. However,
even the simplest version in which $w(r)$ is taken to be
density independent and equal to the normalized Mayer function,
$f=1-\exp(-\beta V(r-r'))$, is enough to give a qualitative 
description of the hard-sphere freezing \cite{WDA}.

In this section we explore the consequences of using the DDF
formalism, (\ref{eq:DF}) with approximate forms for $F[\rho]$,
so that the molecular correlations are represented by an 
approximation to the true equilibrium correlations. To carry out
this analysis we study the same systems described in the 
previous section but instead of taking the exact Percus
density functional, (\ref{eq:percus}), we use an
approximate description of the hard rods similar to those
developed for hard spheres. In particular, the WDA for hard
rods with the exact bulk equation of state and the zeroth-order
constant weight function takes
\begin{equation}
\Delta F_{WDA}[\rho(x)] =  
\int_{-\infty}^{\infty} dx \ \phi(\overline \rho(x))  \ \rho(x),
\label{eq:wda1}
\end{equation}
with the exact $\phi(\rho)$ given by (\ref{eq:nonideal}) and
\begin{equation}
\overline \rho(x)=\int_{-\infty}^{\infty} dx' \rho(x') w(|x-x'|)=
\frac{1}{2 \sigma}\int_{x-\sigma}^{x+\sigma} dx'  \rho(x').
\label{eq:wda2}
\end{equation}

  The time evolution equation from the DDF (\ref{eq:DF}) with this 
density functional approximation is similar to (\ref{eq:rho}) but
with the bracket in the second term given by
\begin{equation}
\frac{\delta \Delta F_{wda}[\rho]}{\delta \rho(x)} =
\frac{1}{2} \left( 
\frac{\rho(x+\sigma,t)} {1-\overline \rho(x+\sigma,t)} +
\frac{\rho(x+\sigma,t)} {1-\overline \rho(x,t)}
-\frac{\rho(x-\sigma,t)}{1-\overline \rho(x-\sigma,t)}
-\frac{\rho(x-\sigma,t)}{1-\overline \rho(x,t)} \right).
\label{eq:rhowda}
\end{equation}
Notice that the weighted density $\overline \rho(x)$ 
averages the density profile from $x-\sigma$ to $x+\sigma$,
while the variable $\eta(x)$, in the exact DF, takes 
the average only over half that distance. There is 
also a different combination of the functions evaluated
in (\ref{eq:rho}) and in (\ref{eq:rhowda}), but it is easy
to check that in the low density limit, truncating at
second order in the density, both equations become identical.
This reflects that the WDA includes 
the exact leading term of the molecular correlations
in a density expansion, while the higher-order terms are
approximated. For the equilibrium properties it is known that
this type of zeroth-order WDA overestimates the effects of the
hard-core packing and it becomes more accurate when the 
weight function is allowed to depend on the weighted density.

  The results obtained with this approximate free energy for the
problems A and B in the previous section are qualitatively similar to
those with the exact $F[\rho]$. In the free expansion of a 
dense state the WDA keeps the oscillations in $\rho(x,t)$
for a longer time than the results with (\ref{eq:rho}), 
reflecting the overestimation of the hard core packing,
but the mean square displacements are very similar to those 
in Figure (\ref{fig:uno}).  The equivalent trends are observed
in the problem B; although there is some difference between the exact
and the WDA equilibrium states, the dynamics is rather similar.
We have also checked, in some cases, that the use of a better WDA
approximation, with the density dependence of the weight function
expanded up to second order, gives results closer to those with the
exact $F[\rho]$. In this way we have checked that the practical use
of the DDF formalism does not depend on having the exact equilibrium
$F[\rho]$. The use of WDA or other density functional approximations
opens a wide road for the study of the dynamical properties
of many systems, with a level of approximation comparable to the
accuracy of the results for the equilibrium properties.

 The system studied in the last subsection provided the most severe test
for the DDF, since it was the case in which the 
molecular correlations played the most important role, and this system
is the one in which the
the use of an approximated $F[\rho]$ may lead to qualitative
differences. We have solved (\ref{eq:DF}) with 
(\ref{eq:rhowda}) for the same system
of $N=4$ hard rods in a periodic external potential as before
and look for the relaxation from the asymmetric density 
distribution. For the system presented in Figure (\ref{fig:cuatro}),
with $\sigma=1.6$ and $\beta V_o=2.$, there is a qualitative difference 
with the results of using the exact $F[\rho]$. With the WDA free energy 
the time evolution of $\rho(x,t)$ goes to a stable  very 
asymmetric profile, with an order parameter $\xi=0.99216$, indicating
that the particles remain mainly in the potential wells where they
are initially located. The freezing of the initial asymmetry is only
possible because the WDA equilibrium free energy for this system
has two different minima, with positive and negative values of $\xi$,
instead of the single minimum with $\xi=0$ given by the exact $F[\rho]$.
The approximate inclusion of the hard-core correlations in the WDA
is not good enough to average over the two types of configurations
which appear as separated equilibrium states and, for the same reason,
the DDF equation cannot include the dynamic path between the two
types of configurations. 

\begin{figure}
\centerline{
\psfig{figure=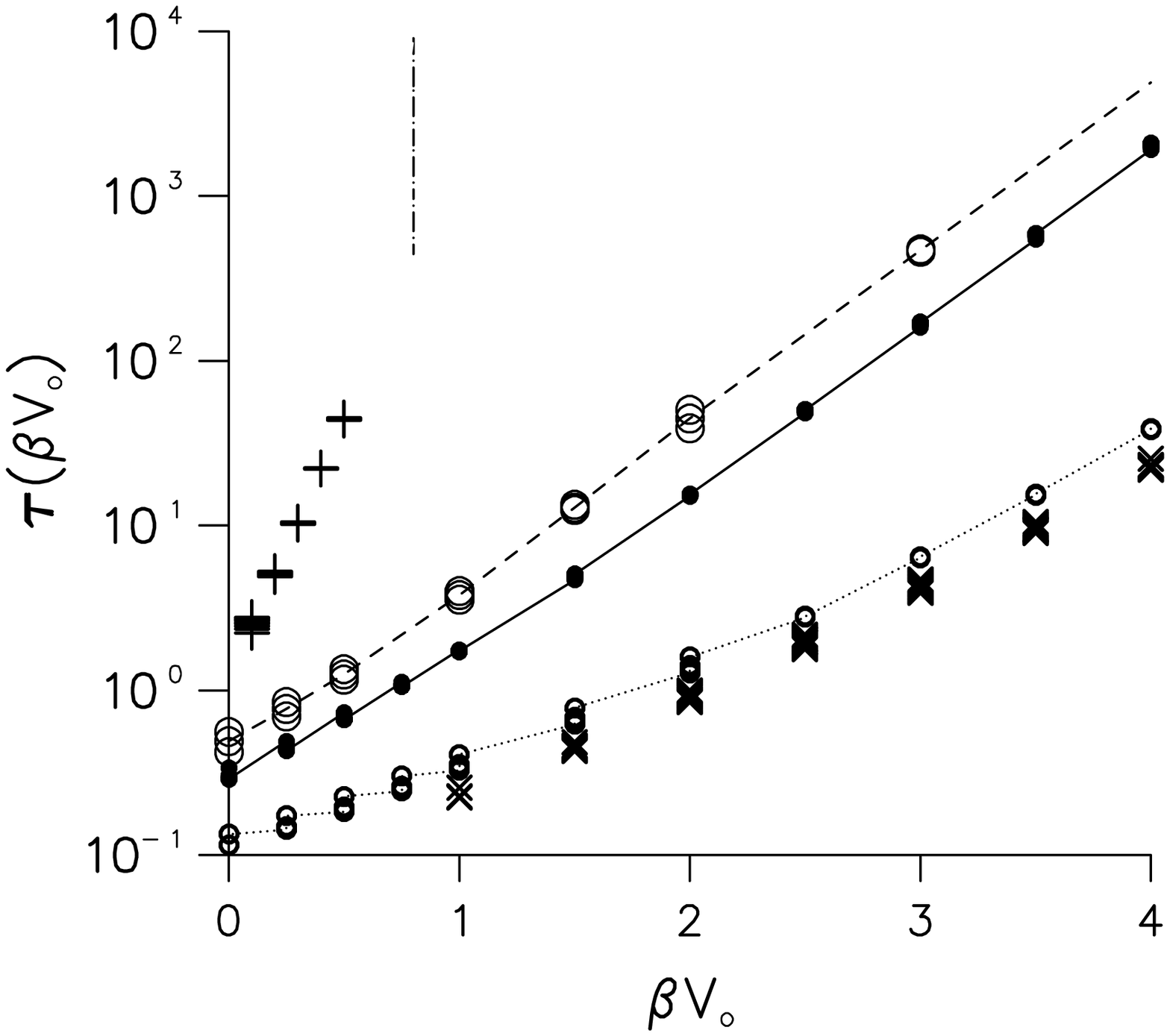,width=10.cm,height=10.cm,angle=-90}}
\caption{Relaxation time $\tau$ for the system in Figure (4) as
a function of the external potential amplitude $\beta V_o$, for
fixed rod length $\sigma=1.6$, 
from the results of the DDF equation with the exact $F[\rho]$
(dots and full line) and the average over 
2000 Langevin  simulations (large circles and dashed line). For comparison 
we include the ideal gas results (small circles and dotted line). In
all the cases the lines are a guide to the eye. We also include the 
results of the DDF equation with approximate $F[\rho]$: WDA ($+$)
and RY ($\times$). The vertical dash-dotted line is the approximated location
of the divergence of the WDA relaxation times, for larger values of
$\beta V_o$ the results of this approximation do not decay to $\xi=0$.
}
\label{fig:seis}
\end{figure}

 For lower values of the external potential
(or higher temperature) we find that the final
order parameter predicted by the WDA decreases, as in a 
phase transition. Indeed, the splitting of the equilibrium state
is nothing but a spurious phase transition predicted by an approximate
$F[\rho]$.  Below a critical value of $\beta V_o\approx 0.7$ the 
WDA recovers a single symmetric equilibrium state with $\xi=0$.
The relaxation time diverges as we approach the phase transition
and it is presented in Figure (\ref{fig:seis}) together with the 
results of the exact $F[\rho]$. This divergence is
obviously a spurious result of the approximation used for $F[\rho]$.
The only reasonable interpretation is that whenever
an approximate free energy density functional has different
local minima, in which the DDF equation may get trapped,
the exact dynamics of the system would require rather long
times to relax the density along the the functional path
related to the 'order parameter' of the spurious phase 
transition.  The DDF with the WDA is good enough to identify
the slow density path, along the parameter $\xi$, and it is
good enough to describe the relaxation dynamics along other
functional directions, but it cannot be used to describe the
relaxation of $\xi$.

  We have checked that on improving the WDA, by including the
density dependence of the weight function, the spurious
phase transition is shifted towards larger values of $\beta V_o$. 
Again a better approximation for the equilibrium free energy
functional also gives a better account of the molecular correlations
in the DDF equation, since 
it is able to find relaxation paths which 'were not seen'
by  simpler DF approximations.  However, when $\beta V_o$
goes beyond a new threshold value (i.e. when the exact relaxation time
becomes too large for the approximate $F[\rho]$) the spurious
phase transition reappears.

 We have also tested the predictions
of the DDF with an even simpler approximation for the equilibrium 
density functional. This is the Ramakrishnan-Youssouf functional
\cite{Dasgupta},\cite{Kirkpatrick},
sometimes referred as the HNC density functional approximation,
which considers a functional Taylor expansion of the free energy about
a global mean density $\rho_b$. Truncating the series 
at the second order (higher order coefficients are in general not known)
one gets:

\begin{equation}
\Delta F_{RY}[\rho] = -\frac{1}{2}\int d {\bf r}\int d {\bf r'}
(\rho({\bf r})-\rho_b) c_b({\bf r-r'})(\rho({\bf r'})-\rho_b),
\label{eq:RY}
\end{equation}
where $c_b({\bf r-r'})$ is the direct
correlation function of a uniform fluid with density
$\rho_b$. The merit of the RY functional is that it is perhaps
the simplest recipe and is easy to implement numerically. However
it requires that the system have a well-defined mean density,
$\rho_b$, and it may  
it become unreliable when the density varies rapidly over
a length scale of the order of molecular size. Notice that
in our case using
the RY approximation is equivalent to saying that during its evolution
the system has the same two point correlations that it would have
at constant density $\rho_b$. 

 In our one-dimensional hard-rod 
system we may use the exact direct correlation function of the 
bulk hard-rods system, and the DDF equation is again similar to
(\ref{eq:rho}) but with the second term bracket given by
\begin{equation}
\frac{\delta \Delta F_{RY}[\rho]} {\delta \rho(x)}=
\left(\frac{\rho(x+\sigma,t)-\rho(x-\sigma,t)}
 {1-\eta_b}+ \rho_b 
\frac {\eta(x+\sigma/2,t)- \eta(x-\sigma/2,t)}{1-\eta_b^2}
\right).
\label{eq:rhory}
\end{equation}

The application to examples A and B is uncertain
because they do not have a clearly defined mean density.
In the third example 
we have used (\ref{eq:DF},\ref{eq:rhory}) with the obvious
choice $\rho_b=N/L=0.5$ to get the relaxation of the 
density distribution in the periodic 
external potential. The results, in Figure (\ref{fig:seis}),
show a surprisingly fast relaxation towards the symmetric 
density profile. This density functional
approximation never has spurious minima, so that the dynamics 
is never trapped in asymmetric distributions, but the relaxation
is even faster than in the ideal gas case, in clear contradiction
with the Langevin simulation. 
This shows an example opposite to the WDA result, the effect of the
approximation for the free energy may be to underestimate
the packing effects. In this case there is no risk that 
the system gets trapped out of the true equilibrium, but the
approximation fails to describe the long relaxation times
imposed by these packing constraints. 

\section{conclusions}

We have presented a dynamic density functional approach for
the relaxation of a classical system in terms of
its equilibrium free energy density functional
$F[\rho]$. The approach
is valid only when the velocity correlations (not included in
$F[\rho]$) are irrelevant, which excludes hydrodynamic modes
and temperature gradients
but may include problems like the Brownian motion of colloidal 
particles and the molecular rearrangements of highly packed
systems, when the collision time is much shorter than the
relaxation time. In the derivation of the approach we 
start with the stochastic equations for the Langevin dynamics
of Brownian particles and get a deterministic DDF equation
for the time dependence of the density distribution, $\rho({\bf r},t)$,
which has to be interpreted as the ensemble average over the 
realizations of the random noise in the Langevin dynamics.
The equilibrium DF formalism is used to go from a formal
BBGKY hierarchy, for the coupled dynamics of the n-particle
distribution functions, to a closed DDF equation for 
$\rho({\bf r},t)$. The assumption leading to this result is that
the correlation structure in the system out of equilibrium is
replaced by that in the equilibrium system with the same 
density distribution.  Moreover, we do not need to calculate
the correlation structure explicitly because its effects on the
dynamics are given directly in terms of the functional derivative
of $F[\rho]$. 

   The main advantage of the approach, for practical purposes, is 
being able to use the good approximations developed for the free energy 
functional of hard-core molecules to include the packing
constraints in the dynamics of dense systems. We have presented
several examples of one-dimensional hard rods to compare
the DDF with the average of Langevin simulations.
The results, with the exact $F[\rho]$, are always in qualitatively 
good agreement, even in those cases
in which the dynamics becomes extremely slow, when the 
relaxation requires very unlikely correlations between the
particles. The main source of discrepancy is probably due to the
fact that the free energy density functional developed for
equilibrium always refers to the grand-canonical ensemble,
so that the DDF includes the relaxation through changes in the
number of particles, while the Langevin dynamics keeps $N$ fixed.
With the use of approximate density functionals, of the same type
as developed for realistic models in three dimensions, we 
get an approximate description of the relaxation dynamics 
of a quality comparable to that for the equilibrium properties
arising from the same $F[\rho]$. In some cases, when the role of the
correlations is not too important and the relaxation times are not
too long, the approximate $F[\rho]$ is fairly accurate for the 
dynamic properties. In the third case, in which strong effects of the
hard core packing lead to very long relaxation times, the 
difference between the results of the exact and the approximate 
$F[\rho]$ are qualitative both for the equilibrium and for the dynamic
properties. The presence of different local minima in the equilibrium
free energy density functional produces, in the approximate DDF
equation, the permanent freezing of system in any of these states,
in contrast with the results of the exact free energy functional.

In an interesting series of papers Kawasaki \cite{Kawasaki}
derived a dynamic equation, in terms of a density functional
hamiltonian $H[\rho]$, by a method different from ours, but his 
resulting Fokker-Planck equation
for the probability distribution of the density:
\begin{equation}
\frac{\partial P[\rho({\bf r}),t]}{\partial t}=
-\int d{\bf r}\frac{\delta}{\delta
\rho({\bf r})}{\bf  \nabla}\cdot\rho({\bf r}) {\bf \nabla}
\bigl[T\frac{\delta}{\delta\rho({\bf r})}+\frac{\delta H[\rho]}{\delta
\rho({\bf r})}\bigr] P[\rho({\bf r}),t] 
\label{eq:FPE}
\end{equation}
is equivalent to our eq. (\ref{eq:DF}), apart from the presence of a 
term due to the stochastic noise. If $\rho$ in Kawasaki equation 
(\ref{eq:FPE}) is interpreted as a density operator, $\hat\rho({\bf
r},t)$, his approach is
equivalent to Dean's equation (\ref{eq:otto}); leading to a 
stochastic equation for this density operator. The main qualitative
difference of this approach with our deterministic 
equation (\ref{eq:DF}) is that, 
when the approximate density functional for the equilibrium free 
energy has different local minima, the random noise in the DDF
equation would always give a chance for changing from one minimum to 
another. The long time average of the density would always
be a superposition of the density in the different local minima
and the long relaxation times would appear as the result of
high barriers between the local minima.  All these features may 
appear to be physically correct and to represent a qualitative
improvement over the deterministic DDF developed here. However,
the use of the equilibrium free energy density functional
requires always a density defined as a thermal ensemble
average, while (\ref{eq:otto}) refers to the instantaneous
density operator. The hamiltonian $H$ in (\ref{eq:FPE}) should 
also be a functional of the density operator, $\hat\rho({\bf r},t)$,
and it is a completely different mathematical object that the
excess free energy $\Delta F[\rho]$, as a functional of the
equilibrium density $\rho_o({\bf r})$. Thus, equations
(\ref{eq:otto}) and (\ref{eq:FPE}) are correct, for the
density operator $\hat\rho({\bf r},t)$ but impossible to
translate in terms of the equilibrium free energy density functional.
On the other hand, if these equations are interpreted as
equations for the ensemble averaged density, $\rho({\bf r},t)$,
the random noise term would 
lead to a double counting of thermal fluctuations
in the equation for $\rho({\bf r},t)$. In particular, it would lead
to a wrong equilibrium distribution, as can be verified in the simple
case of non interacting particles for which the FPE (\ref{eq:FPE}) 
converges to a probability distribution:
\begin{equation}
P_{eq}[\rho(r)]\sim \exp(-F[\rho]/k_B T)=
\exp \left[ -\int d r \ \rho({\bf r}) \ (\ln \rho({\bf r})-1)+
\beta V_{ext}({\bf r}) \right],
\label{eq:false}
\end{equation}
where the first term in the exponential is clearly due to the overcounting of  
thermal fluctuations.

   Moreover, we have shown that if we use the exact functional
$F[\rho]$ the DDF equation does not require random noise to give
the correct results: the system always flows towards the true and
unique equilibrium state. The DDF relaxation time may become very long
when the system has to go through highly correlated states, but
this effect corresponds to systems in which the true relaxation time 
(in the Langevin description) is also very long. When we use
an approximate $F[\rho]$ these long relaxation times may become 
infinite, with the system trapped at a local minima of
$F[\rho]$. An attempt to avoid this effect would require 
the intensity of random noise not to be proportional to the
temperature, but to the error made by the approximation to 
$F[\rho]$, an error which is obviously unknown until we make a 
better approximation. Otherwise, the relaxation time given by the
stochastic DDF in these cases would just be a direct result
of the uncontrolled level of noise kept in the functional 
equation.

  Within our deterministic DDF, the existence of frozen 
states in the local minima of the approximate $F[\rho]$
should be interpreted as the signature of very long
relaxation times, but the only way to calculate how long
these times are is to improve the approximation 
for the equilibrium free energy. Nevertheless, knowing
the existence and the approximate structure of these
states, as given by workable approximations for $F[\rho]$,
is already an interesting use of the DDF, together with 
its use to study the relaxation process in those cases
in which there are no problems with different local 
minima.

  Finally, as a plan for future work, we can consider  
a systematic way to improve the use of the equilibrium
$F[\rho]$ to estimate the correlation structure in the following
terms: We tag particle number 1 and
follow its position ${\bf r}_1(t)$ separately, while 
all the other particles  ($i=2,..,N$) are included in a density 
description, with $\rho({\bf r}_1(t),{\bf r},t)$ as the
noise-averaged conditional probability of finding a
particle at position ${\bf r}$, and time $t$, if the
tagged particle is at position ${\bf r}_1$. Now, we may
consider that the $N-1$ particles are moving in an 
effective external potential, $V_{ext}({\bf r})+ V({\bf r}-{\bf r}_1)$,
which also contains the interaction with the tagged particle.
The equivalent to  the DDF equation (\ref{eq:DF}) may be applied to
the conditional density $\rho({\bf r}_1,{\bf r},t)$,
which is now coupled to the stochastic equation for 
${\bf r}_1(t)$. The advantage is that the correlation structure 
becomes partially described at the level of an effective 
one-particle density in an effective external potential,
which is in principle much easier to describe with approximate 
free energy density functionals. This type of description would be 
similar to the 'reaction path' description of chemical reactions,
in which one, or a few, variables are used to describe the
relevant functional directions for the changes in the molecular
conformations. However, the formal and the practical use
of density functional approximations along this line is still 
an open problem. 

\section{Acknowledgments}
 It is a pleasure to thank P. Tartaglia and F. Sciortino for many
illuminating discussions, and J.P. Hernandez for carefull reading of the
manuscript.  P.T acknowledges financial support
from the Direcci{\'o}n General de Ense\~nanza
Superior e Investigaci{\'o}n Cient{\'{\i}}fica of Spain under 
grant number PB97-1223-C02-01. UMBM was partially supported by 
Istituto Nazionale di Fisica Nucleare, project FI11.

\newpage

\begin{references}
\bibitem{footnote} The derivation of the equation 
for the averages could have been performed
by employing the Stratonovich convention with the same result.
\end{references}
\end{document}